\documentclass{jfm}

\usepackage[latin1]{inputenc}
\usepackage{amsmath}    
\usepackage{amssymb}
\usepackage{graphicx}   
\usepackage{verbatim}   
\usepackage{color}      
\usepackage{subfigure}  
\usepackage{hyperref}   
\usepackage{tensor}
\usepackage{natbib}

\newcommand{\bnabla}{\boldsymbol{\nabla}}

\newcommand{\bOmega}{\boldsymbol{\Omega}}

\newcommand{\la}{\left<}
\newcommand{\ra}{\right>}
\newcommand{\tu}{\tilde{u}}
\newcommand{\tv}{\tilde{v}}
\newcommand{\tw}{\tilde{w}}

\newcommand{\Bx}{{\cal B}_x}
\newcommand{\By}{{\cal B}_y}
\newcommand{\hbB}{\hat{{\bf B}}}
\newcommand{\hbv}{\hat{{\bf v}}}
\newcommand{\cB}{{\boldsymbol{\cal B}}}
\newcommand{\Rm}{\text{Rm}}
\newcommand{\Pm}{\text{Pm}}

\newcommand{\bcU}{\boldsymbol{\cal U}}

\newcommand{\cor}[1]{{#1}}

\title[Dynamo saturation down to vanishing viscosity]{Dynamo saturation down to vanishing viscosity: strong-field and inertial scaling regimes.}
\author{Kannabiran Seshasayanan, Basile Gallet}
\affiliation{Service de Physique de l'\'Etat Condens\'e, CNRS UMR 3680, CEA Saclay, 91191 Gif-sur-Yvette, France}

\begin{document}

\maketitle

\begin{abstract}
We present analytical examples of fluid dynamos that saturate through the action of the Coriolis and inertial terms of the Navier-Stokes equation. The flow is driven by a body force and is subject to global rotation and uniform sweeping velocity. The model can be studied down to arbitrarily low viscosity and naturally leads to the strong-field scaling regime for the magnetic energy produced above threshold: the magnetic energy is proportional to the global rotation rate and independent of the viscosity $\nu$. Depending on the relative orientations of global rotation and large-scale sweeping, the dynamo bifurcation is either supercritical or subcritical. In the supercritical case, the magnetic energy follows the scaling-law for supercritical strong-field dynamos predicted on dimensional grounds by P\'etr\'elis \& Fauve (2001). In the subcritical case, the system jumps to a finite-amplitude dynamo branch. \cor{The magnetic energy obeys a magneto-geostrophic scaling-law \citep{PHRoberts72}, with a turbulent Elsasser number of the order of unity, where the magnetic diffusivity of the standard Elsasser number appears to be replaced by an eddy diffusivity.} In the absence of global rotation, the dynamo bifurcation is subcritical and the saturated magnetic energy obeys the equipartition scaling regime. We consider both the vicinity of the dynamo threshold and the limit of large distance from threshold to put these various scaling behaviors on firm analytical ground.
\end{abstract}

\cor{A key challenge in dynamo theory is to predict the strength of the generated magnetic field. This is of obvious interest in an astrophysical context, where one would like to estimate the magnetic fields of astrophysical objects, but also in the context of laboratory experiments, where many questions arise regarding the saturation mechanisms of instabilities arising over high-Reynolds-number background flows \citep{Petrelis2007,Gallet2012,Fauve}.}
\cor{Predictions of the intensity of the dynamo field mainly rely on dimensional analysis,} and the resulting scaling-laws depend on the dominant balance at stake in the Navier-Stokes equation: in simple viscous analytical models, the main balance is between the Lorentz force and the viscous one, the magnetic energy being proportional to molecular viscosity \citep{Soward,Gilbert,Nunez}. In contrast with such viscous models, natural dynamos in planets and stars as well as laboratory experiments operate at large Reynolds number. For such laboratory flows \citep{Gailitis,Stieglitz,Monchaux}, the dominant balance is between the Lorentz force and the advective term. This leads to an ``inertial'' or ``turbulent'' scaling regime where the magnetic energy is independent of molecular viscosity \citep{Petrelis2001,Petrelis2007,Gallet2009}. 

\cor{The situation is even more complex for rapidly rotating flows, where a significant fraction of the Lorentz force can be balanced by the Coriolis term. Motivated by the geodynamo, Roberts put forward a detailed picture of the parameter space of convectively-driven rapidly-rotating dynamos \citep{Roberts78,Roberts88}. He conjectured the coexistence of two dynamo branches. As the Rayleigh number increases, a convective flow sets in above a critical Rayleigh number $Ra_c$. Above some value $Ra_m>Ra_c$ of the Rayleigh number, the flow becomes dynamo-capable and a supercritical branch of dynamo states arises. This supercritical branch is called the ``weak-field'' branch, as analytical examples of such supercritical convective dynamos indicate that the magnetic energy is proportional to the small molecular viscosity on that branch \citep{Soward}. However, based on the linear stability analysis of thermal convection subject to global rotation and uniform external magnetic field, Roberts conjectured the existence of a second dynamo branch coexisting with the weak-field one in parameter space. Indeed, because the combination of global rotation and uniform magnetic field decreases strongly the threshold for convective motion, Roberts argues that this second dynamo branch may appear through a saddle-node bifurcation even below $Ra_c$. For this phenomenon to happen, the magnetic field must be large: it is independent of molecular viscosity and proportional to the global rotation rate \citep{Roberts88,Roberts92}. The corresponding dynamo branch is thus referred to as the ``strong-field'' one. On the strong-field branch, the dominant balance in the Navier-Stokes equation cannot be purely between the Lorentz force and the Coriolis term, because the latter does not do any work. A third force must come into play to provide the energy that is dissipated ohmically. The dominant force balance is thus between the Lorentz force, buoyancy force and Coriolis term. It is often referred to as MAC balance, for Magnetic-Archimedean-Coriolis. While these studies were originally motivated by the geodynamo problem, both the inertial scaling regime described above and the strong-field branch have been argued to bear some relevance to stellar magnetic fields \citep{Morin}.}

Testing these predictions in fully 3D direct numerical simulations (DNS) is extremely challenging, because of the moderate Reynolds-number values achievable on modern supercomputers: 3D DNS of the dynamo effect remain strongly influenced by viscous effects \citep{Oruba}. Such numerical studies are therefore in stark contrast with the few successful dynamo experiments, all of which point towards a turbulent saturation regime. The most recent ones clearly point towards a MAC balance in the bulk of the flow \citep{Yadav,Schaeffer}, but they could not establish the independence of magnetic energy with respect to viscosity.
\cor{The problem of clearly identifying a strong-field dynamo branch remains overwhelming, and efforts have therefore split into two kinds of studies:
\begin{enumerate}
\item Some studies retain the full complexity of the convective dynamo problem, and aim at reproducing the multiple-branch picture conjectured by Roberts. These studies either consider the full set of convective MHD equations, or focus on precise asymptotic limits to derive reduced sets of equations that can be simulated at lower computational cost \citep{Calkins,Calkins2016,Plumley}. In this quest for numerically tractable asymptotic regimes, another line of work focused on the rapid-rotation limit in otherwise viscous flows \citep{Hughes,Cattaneo17,Dormy}. The former two studies clearly evidenced dynamo states for which the Lorentz force contributes to the dominant force balance. This results in very different velocity fields during the kinematic and dynamic phases of dynamo action. The study by \citet{Dormy} clearly showed the coexistence of branches of weaker and stronger magnetic field. However, in spite of the qualitative agreement with Roberts' picture, the agreement cannot be made quantitative as viscosity still plays a central role in setting the magnetic field strength.
\item Another approach to the problem consists in replacing the complex convective flow by a body-forced one (see e.g. \citet{Moffatt}) or even a boundary-driven one \citep{Petrelis2001}. Such driving mechanisms are arguably more relevant to  dynamo experiments, but most importantly they allow for simpler analytical treatment. The goal here is to reproduce the viscosity-independent scaling behaviour of the magnetic energy at high Reynolds number. 
As for convective systems, the body-forced equations can be studied in their full complexity, or using reduced sets of asymptotic equations. As an example, we recently derived a reduced set of quasi-2D equations that is asymptotically valid in the limit of rapid rotation and in the vicinity of the dynamo threshold. We could then simulate these reduced equations down to very low values of the magnetic Prandtl number, thereby showing that the magnetic energy transitions to the turbulent scaling regime for low enough magnetic Prandtl number, $\Pm \lesssim 10^{-3}$ \citep{Seshasayanan}.
\end{enumerate}}

\cor{To summarize, studies of type a) can recover the multiple-branch picture but cannot achieve the viscosity-free scaling-laws of a strong-field dynamo, whereas studies of type b) successfully realize these viscosity-free scaling-laws, but cannot produce Roberts' multiple-branch picture. By extension, in several previous studies of type b) \citep{Petrelis2001}, as well as in the present one, the ``strong-field scaling regime'' then refers to a dynamo branch where the magnetic energy is independent of molecular viscosity and proportional to the global rotation rate, regardless of whether the branch is subcritical or supercritical, and whether it coexists with a weaker-field branch or not. In the context of laboratory experiments, studies of type b) can be relevant as such, while in the context of astrophysical dynamos the hope is that the qualitative picture arising in studies of type a) can be combined with the quantitative scaling-laws arising in studies of type b).}

\cor{The following study belongs to type b) above: in the present context of limited numerical evidence, simple analytical examples of dynamos displaying the strong-field or turbulent scaling-regimes are highly desirable. We thus introduce body-forced flows for which the dynamo saturation can be studied analytically down to arbitrarily low viscosity. It is based on the standard G.O. Roberts flow, to which we add two additional ingredients: in the presence of global rotation and/or uniform large-scale sweeping flow, we show that the dynamo instability saturates through the action of the Coriolis and/or inertial terms of the Navier-Stokes equation. The resulting magnetic energy is independent of molecular viscosity when the latter is low enough, and it is proportional to the global rotation rate for rapid rotation. It therefore reproduces the scaling behaviour of a strong-field dynamo branch, without the full complexity of thermal convection. Because mechanical forcing replaces the buoyancy force, the standard MAC balance is replaced by a Magnetic-Forcing-Coriolis balance, or MFC if one insists on using acronyms. This MFC balance arises directly at the beginning of the dynamo branch when the bifurcation is subcritical, and at large distance from threshold when the bifurcation is supercritical.}

Our approach is fully nonlinear and relies on scale separation only: the lengthscale of the flow is much less than that of the magnetic field. We therefore relax the common assumption of weak departure from the dynamo threshold, which provides an analytical avenue to study the magnetic energy produced far from threshold. The resulting dynamo branch depends on whether global rotation is present:
\begin{itemize}
\item Without global rotation but in the presence of a large-scale sweeping flow, the dynamo instability is subcritical. The magnetic energy is independent of viscosity and corresponds to a regime of equipartition between kinetic and magnetic energy. It is therefore independent of both viscosity and magnetic diffusivity.
\item When both global rotation and large-scale sweeping flow are present, the nature of the dynamo bifurcation (supercritical or subcritical) depends on their relative orientations and strengths. The magnetic energy is proportional to the global rotation rate and independent of viscosity, therefore achieving the strong-field scaling regime. Far away from threshold, it is also independent of the magnetic diffusivity: \cor{the ratio of kinetic to magnetic energy is then simply given by the Rossby number, which corresponds to the ``magneto-geostrophic'' scaling-law proposed in \citet{PHRoberts72}. }
\end{itemize}

In section \ref{secthsetup} we introduce the theoretical setup and derive the nonlinear $\alpha$-effect in the presence of sweeping flow, background rotation, viscosity, and magnetic feedback through the Lorentz force. In section \ref{seclinstabWNL} we present the linear stability and weakly nonlinear analyses of the resulting equations for the large-scale magnetic field. We derive the scaling behaviour for the magnetic energy in the vicinity of the dynamo bifurcation. We then study the dynamo branches at arbitrary distance from threshold in section \ref{secfinite}, establishing regimes where the magnetic energy is independent of both viscosity and magnetic diffusivity. Section \ref{secdisc} is devoted to a discussion of three important points: the case of a large-scale zonal flow, the stability of the present analytical solutions and the criteria to achieve viscosity-independent scaling regimes in dynamo simulations (low Ekman number versus low-magnetic Prandtl number).

\section{Theoretical setup \label{secthsetup}}

\subsection{Body-forced flow subject to sweeping and rotation}

\begin{figure}
    \centerline{\includegraphics[width=6 cm]{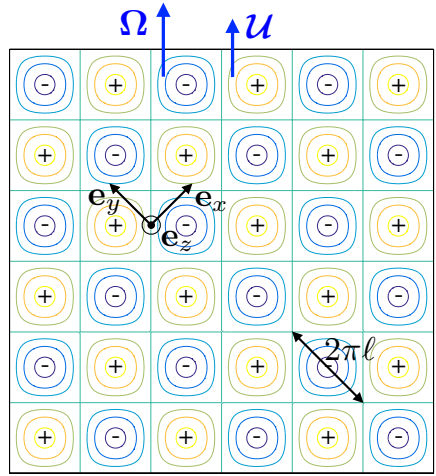} }
   \caption{An electrically conducting fluid is stirred by a steady body-force. We represent the projection of the force-field lines in an $(x,y)$ plane, together with the sign of the $z$ component. Light-color lines rotate counterclockwise, while dark-colored lines rotate clockwise. The flow is subject to global rotation with rotation vector $\bOmega$, and advection by a uniform sweeping velocity $\bcU$. We describe in detail the situation where $\bOmega$ and $\bcU$ are collinear, the case of a zonal flow $\bcU$ perpendicular to $\bOmega$ being left for the discussion section.  \label{schema}}
\end{figure}

We consider an electrically conducting Newtonian fluid of density $\rho$ and kinematic viscosity $\nu$ inside a cubic periodic domain of sidelength $\lambda$. A steady body-force ${\bf F}^*$ drives a small-scale flow of G.O. Roberts geometry \citep{Roberts72}:
\begin{equation}
{\bf F}^* = \frac{F^*}{2} \left\{ \begin{matrix}
 e^{i y^*/\ell} \\
 e^{i x^*/\ell} \\
i \, e^{i x^*/\ell} - i \, e^{i y^*/\ell}
\end{matrix} \right. + c.c. \, ,\label{forceF}
\end{equation}
where $(x^*,y^*,z^*)$ denote standard Cartesian coordinates, $\ell \ll \lambda$, and $F^*>0$. Such small-scale body forces are routinely used to drive helical flows in dynamo studies \citep{Moffatt}. At larger scale, mechanical forcing is becoming increasingly popular as an alternate driving mechanism of some astrophysical dynamos \citep{LeBars}. The cubic domain lies in a frame rotating with a rotation vector $\bOmega=\omega/2 \, (\textbf{e}_x + \textbf{e}_y)$. Additionally, we consider the presence of a uniform time-independent sweeping velocity $\boldsymbol{\cal U} = {\cal U}  (\textbf{e}_x + \textbf{e}_y)$. Such a sweeping flow can be specified at the outset, just as we specify global rotation: because of momentum conservation, the uniform velocity $\boldsymbol{\cal U}$ is unaffected by the smaller-scale flow driven by the mean-zero field ${\bf F}^*$. The total velocity field therefore reads ${\bf u}=\boldsymbol{\cal U}+{\bf v}^*$, where ${\bf v}^*$ denotes the small-scale velocity field driven by ${\bf F}$. We focus on the situation where the large-scale sweeping flow is parallel to the direction of global rotation ; the case of a zonal flow, perpendicular to $\bOmega$, is left for the discussion section \ref{secsweeping}. A similar combination of cellular flow and sweeping velocity was considered by \citet{Tilgner}, who focuses on the kinematic dynamo problem for small-scale dynamo modes in the absence of global rotation. By contrast, the present study focuses on fully nonlinear dynamos operating in the limit of scale separation $\ell\ll \lambda$.

We non-dimensionalize the equations using the length scale $\ell$ and the time scale $\ell^2/\eta$, where $\eta=1/\mu_0 \sigma$ is the magnetic diffusivity, with $\mu_0$ the magnetic permeability of vacuum and $\sigma$ the electrical conductivity of the fluid. We denote as ${\bf B}^*$ the (dimensional) magnetic field. We introduce the dimensionless variables:
\begin{eqnarray}
{\bf x} = \frac{{\bf x}^*}{\ell} \, , \qquad   t=\frac{t^* \eta}{ \ell^2} \, , \qquad       {\bf v} = \frac{{\bf v}^* \ell}{\eta} \, , \qquad  {\bf F} = \frac{{\bf F}^* \ell^3}{\eta^2} \, , \qquad {\bf B} = \frac{{\bf B}^* \ell}{\sqrt{\rho \mu_0} \eta} \, ,
\end{eqnarray}
where the quantities with a $^*$ are dimensional, while the quantities without a $^*$ are their dimensionless counterparts.
In terms of the dimensionless variables, the Navier-Stokes and induction equations read:
\begin{eqnarray}
\partial_t {\bf v} +   R [({\bf e}_x  + {\bf e}_y) \cdot \bnabla] {\bf v} + & & \frac{R}{Ro} ({\bf e}_x + {\bf e}_y) \times {\bf v}  + ({\bf v} \cdot \bnabla) {\bf v} \label{dimlessNS}\\
\nonumber & & = -\bnabla p + \Pm \, \bnabla^2 {\bf v} + ({\bf B} \cdot \bnabla) {\bf B} + {\bf F} \, , \\
\partial_t {\bf B} + R [({\bf e}_x + {\bf e}_y) \cdot \bnabla] {\bf B} & & = \bnabla \times ({\bf v} \times {\bf B}) + \bnabla^2 {\bf B} \, , \label{dimlessInd}
\end{eqnarray}
where $p$ is the generalized pressure, and we have introduced the following dimensionless parameters:
\begin{eqnarray}
R=\frac{{\cal U} \ell}{\eta} \, , \qquad  Ro=\frac{{\cal U} }{\ell \omega}  \, ,  \qquad  \Pm=\frac{\nu}{\eta} \, ,
\end{eqnarray}
$\nu$ being the kinematic viscosity of the fluid. $R$ is a magnetic Reynolds number built with the sweeping flow -- it is also the dimensionless sweeping velocity -- and $Ro$ is a Rossby number built with the sweeping flow, the global rotation rate and the forcing scale.
The second term on the left-hand side of equations (\ref{dimlessNS}-\ref{dimlessInd}) corresponds to advection by the sweeping flow $\bcU$, while the third term of the Navier-Stokes equation (\ref{dimlessNS}) is the Coriolis force associated to $\bOmega$. The flow being incompressible, equations (\ref{dimlessNS}-\ref{dimlessInd}) are supplemented by the divergence-free constraints:
\begin{eqnarray}
\bnabla \cdot {\bf v} = 0 \, \qquad \text{and} \qquad \bnabla \cdot {\bf B} = 0 \, .
\end{eqnarray}



\subsection{Scale-separation}

We follow the standard procedure of the mean-field dynamo framework, making use of scale separation: the scale $\ell$ of the forcing is much smaller than the extension $\lambda$ of the domain, and we define the small parameter $\epsilon=\ell / \lambda \ll 1$. We introduce a slow timescale, together with a slowly varying vertical coordinate:
\begin{eqnarray}
T = \epsilon^2 \,  t \, , \qquad Z=\epsilon \,  z \, .
\end{eqnarray}
We consider the following scalings for the small-scale flow, large-scale flow, global rotation rate and forcing:
\begin{eqnarray}
{\bf v}= {\cal O}(\sqrt{\epsilon}) \, , \qquad R= {\cal O}(1) \, , \qquad Ro={\cal O}(1) , \qquad F={\cal O}(\sqrt{\epsilon}) \, .
\end{eqnarray}
In appendix \ref{appeq}, we reproduce the standard multiple-scale expansion leading to the concept of $\alpha$-effect, with the addition of global rotation and large-scale sweeping flow: the magnetic field ${\bf B}$ decomposes into an ${\cal O}(1)$ large-scale magnetic field ${\cB}(Z,T)$ that depends on the slow time and space variables only, together with a weaker $O(\sqrt{\epsilon})$ small-scale magnetic field ${\bf b}(x,y,Z,t,T)$ that depends on both fast and slow coordinates. These two fields obey the following set of equations:
\begin{eqnarray}
\partial_t {\bf b} + R [({\bf e}_x + {\bf e}_y) \cdot \bnabla_{\bf x}] {\bf b} & = & ({\cB} \cdot \bnabla_{\bf x}) {\bf v} + \bnabla_{\bf x}^2 {\bf b} \, , \label{ssind}\\
\partial_T {\cB} & = & \epsilon^{-1} \bnabla_{\bf X} \times \la {\bf v} \times {\bf b} \ra + \bnabla_{\bf X}^2 {\cB} \, , \label{lsind}
\end{eqnarray}
where $\bnabla_{\bf x}=(\partial_x, \partial_y, 0)$, $\bnabla_{\bf X}=(0,0,\partial_Z)$, and $\la \cdot \ra$ denotes an average over the fast variables $x$, $y$ and $t$. \cor{Even though there is a factor $\epsilon^{-1}$ in the first term on the right-hand side of equation (\ref{lsind}), we stress the fact that all the terms of this equation arise at the same order in the asymptotic expansion for the appropriately-scaled fields (see details in Appendix A).} The incompressibility constraint yields $\bnabla_{\bf x} \cdot {\bf b} = 0$ and $\bnabla_{\bf X} \cdot {\cB} = 0$. These induction equations are supplemented by an equation governing the evolution of the small-scale velocity field:
\begin{eqnarray}
\partial_t {\bf v} +   R [({\bf e}_x  + {\bf e}_y) \cdot \bnabla_{\bf x}] {\bf v} + & & \frac{R}{Ro} ({\bf e}_x + {\bf e}_y) \times {\bf v}  \label{reducedNS}\\
\nonumber & & = -\bnabla_{\bf x} p + \Pm \, \bnabla_{\bf x}^2 {\bf v} + ({\cB} \cdot \bnabla_{\bf x}) {\bf b} + {\bf F} \, ,
\end{eqnarray}
together with the incompressibility constraint $\bnabla_{\bf x} \cdot {\bf v} = 0$. An important outcome of this approach is that the nonlinearity $({\bf v} \cdot \bnabla) {\bf v}$ of the Navier-Stokes equation (\ref{dimlessNS}) is subdominant and does not appear in (\ref{reducedNS}). The set of equations (\ref{ssind})-(\ref{reducedNS}) is a closed set of equations from which one can compute the dynamo branches of the system. 

\subsection{Solution for the small-scale fields}

The first step of the dynamo computation consists in assuming the existence of a large-scale field $\cB$. Because $\bnabla_{\bf X} \cdot {\cB} = 0$, this field has no component along $z$ and we write $\cB=(\Bx,\By,0)$. Substitution into (\ref{ssind}) and (\ref{reducedNS}) leads to a set of linear equations for the small-scale fields ${\bf b}$ and ${\bf v}$. Neglecting the short transient, we focus on the $t$-independent solutions to these equations. To eliminate pressure, one can take the curl of the Navier-Stokes equation, $\bnabla_{\bf x} \times$(\ref{reducedNS}), which yields:
\begin{equation}
 R [({\bf e}_x  + {\bf e}_y) \cdot \bnabla_{\bf x}](\bnabla_{\bf x} \times {\bf v}) - \frac{R}{Ro} (\partial_x {\bf v} + \partial_y {\bf v} ) = {\bf F} + \bnabla_{\bf x} \times [({\cB}\cdot \bnabla) {\bf b} ] +\Pm \bnabla_{\bf x}^2 (\bnabla_{\bf x} \times {\bf v})\, , \label{vorticityadim}
\end{equation}
where we used the identity $\bnabla_{\bf x} \times {\bf F}= {\bf F}$. The solution for the small-scale velocity field is of the form:
\begin{equation}
{\bf v} = \left\{ \begin{matrix}
\tu \, e^{i y} \\
\tv \, e^{i x} \\
\tw^{(x)} \, e^{i x} + \tw^{(y)} \, e^{i y}
\end{matrix} \right. \, + c.c. \, , \label{formv}
\end{equation}
where $c.c.$ denotes the complex conjugate, and the coefficients $\tu, \tv, \tw^{(x)}$ and $\tw^{(y)}$ will be determined shortly. They are independent of $x$, $y$ and $t$ but may depend on $Z$ and $T$. Substitution of this form into (\ref{ssind}) leads to the expression of the small-scale magnetic field ${\bf b}$ in terms of ${\bf v}$:
\begin{equation}
{\bf b} = \frac{i }{ 1+i R}  \left\{ \begin{matrix}
\By \tu \, e^{i y} \\
\Bx \tv \, e^{i x} \\
\Bx \tw^{(x)} \, e^{i x} + \By \tw^{(y)} \, e^{i y}
\end{matrix} \right.  \, + c.c. \, .\label{linkbv}
\end{equation}
We finally insert expressions (\ref{formv}) and (\ref{linkbv}) into the vorticity equation (\ref{vorticityadim}) to determine $\tu, \tv, \tw^{(x)}$ and $\tw^{(y)}$:
\begin{eqnarray}
{\tilde u} & = & \frac{F}{2 \left[ i R \left( 1 - Ro^{-1} \right) +\Pm+\frac{\By^2}{(1+iR)}\right]} \, ,\label{expru}\\
{\tilde v} & = & \frac{F}{2 \left[ i R \left( 1 - Ro^{-1} \right) +\Pm+\frac{\Bx^2}{(1+iR)}\right]}  \, , \\
\tw^{(x)} & = & \frac{i F}{2 \left[ i R \left( 1 - Ro^{-1} \right) +\Pm+\frac{\Bx^2}{(1+iR)}\right]}  \, , \\
\tw^{(y)} & = & \frac{- i F}{2 \left[ i R \left( 1 - Ro^{-1} \right) +\Pm+\frac{\By^2}{(1+iR)}\right]}  \, .\label{exprwy}
\end{eqnarray}
This completes the determination of the small-scale velocity and magnetic fields in terms of the large-scale magnetic field $\cB$.

\subsection{$\alpha$-effect and evolution of the large-scale magnetic field}

The goal is now to write a closed equation for the evolution of the large-scale magnetic field $\cB$. From the expressions of the small-scale fields ${\bf v}$ and ${\bf b}$, we compute the mean electromotive force $ \la {\bf v} \times {\bf b} \ra $ appearing in the large-scale induction equation (\ref{lsind}). We obtain:
\begin{equation}
 \la {\bf v} \times {\bf b} \ra = \left\{ \begin{matrix}
 \alpha_{xx} \, \Bx\\
\alpha_{yy} \, \By \\
0
\end{matrix} \right. \, ,
\end{equation}
where the $\alpha$-effect coefficients are:
\begin{eqnarray}
\alpha_{xx;yy} = \frac{-F^2 }{ (1+R^2) \left[ \left( \frac{ {\cal B}_{x;y}^2}{ (1+R^2)}+ \Pm \right)^2 + R^2 \left(1-Ro^{-1}-\frac{ {\cal B}_{x;y}^2 }{ (1+R^2)} \right)^2   \right]} \, .\label{alphaF}
\end{eqnarray}

Dynamo computations are more easily compared to experiments using velocity scales. We therefore introduce the dimensional root-mean small-scale kinetic energy per unit mass of fluid $V^*= \sqrt{\la {\bf v}^{*2}\ra / 2}$. We denote as $\Rm$ the magnetic Reynolds number based on $V^*$:
\begin{eqnarray}
\Rm = \frac{V^* \ell}{\eta} = \sqrt{\frac{\la {\bf v}^2\ra}{2}} \, ,\label{defRm}
\end{eqnarray}
evaluated for the non-magnetic solution. Setting $\Bx=\By=0$ in expressions (\ref{expru}-\ref{exprwy}), we obtain the expression of $F$ in terms of $\Rm$:
\begin{eqnarray}
F = \Rm \, \sqrt{ R^2 \left( 1- Ro^{-1} \right)^2  + \Pm^2} \, ,
\end{eqnarray}
which, after substitution into (\ref{alphaF}), leads to:
\begin{eqnarray}
{\alpha}_{xx;yy} =    -\frac{\Rm^2}{1+R^2} \times  \frac{(Ro^{-1}-1)^2+ \frac{1}{Re^2}}{ \left(\frac{{\cal B}_{x;y}^2 }{R (1+R^2)} + \frac{1}{Re} \right)^2     + \left( Ro^{-1}-1+\frac{{\cal B}_{x;y}^2}{1+R^2 } \right)^2     }      \, ,\label{alphaV}
\end{eqnarray}
where we have introduced the sweeping Reynolds number $Re={\cal U} \ell / \nu$.

The limit of vanishing sweeping flow and global rotation is obtained by taking $(R,Re) \to (0,0)$ with $R=\Pm \, Re$ and fixed $Ro$. In this limit, we recover the standard expression of the viscously-quenched $\alpha$-effect:
\begin{eqnarray}
{\alpha}_{xx;yy} \to    \frac{-\Rm^2}{\left(1+ \frac{{\cal B}_{x;y}^2 }{\Pm}  \right)^2 }      \, ,
\end{eqnarray}
which leads to the viscous regime of dynamo saturation \citep{Gilbert}. The present study focuses on the opposite limit: in the following we show that large-scale sweeping flow and global rotation respectively lead to the inertial and strong-field scaling regimes of dynamo saturation.

\subsection{Neglecting viscous effects \label{secneglectviscous}}

One can readily learn much about the saturation of the dynamo instability by studying the nonlinear $\alpha$-effect coefficients (\ref{alphaV}). Of particular interest is the regime where viscous effects can be neglected. This amounts to neglecting the terms involving the Reynolds number in (\ref{alphaV}). At the numerator and in the ${\cB}$-independent terms of the denominator, the condition to neglect such terms is:
\begin{eqnarray}
\frac{1}{Re} \ll |Ro^{-1}-1| \, .
\end{eqnarray}
Without global rotation, the condition simply becomes $Re \gg 1$ and $Re \gg R^{-1}$. However, for rapid global rotation $|Ro| \ll 1$ the criterion becomes:
\begin{eqnarray}
\frac{\nu}{\ell^2 \omega} \ll 1 \, . \label{criterion1}
\end{eqnarray}
In other words, these viscous contributions can be neglected provided the Ekman number is low enough.
A closer look at the quadratic term in ${\cal B}$ at the denominator of (\ref{alphaV}) gives an additional criterion to neglect viscosity in the quenching of the $\alpha$-effect: in the limit of rapid rotation, the viscous contribution to this term can be neglected provided:
\begin{eqnarray}
\frac{\nu}{\ell^2 \omega} \ll R \, . \label{criterion2}
\end{eqnarray}
Here the Ekman number must be small compared to the magnetic Reynolds number associated to the large-scale flow. 
To summarize, provided the Ekman number is low enough, viscosity can be neglected in the expression (\ref{alphaV}) of the nonlinear $\alpha$-effect, and the dynamo saturation will not involve viscosity. The importance of a low Ekman number was recently highlighted by \cite{Dormy} (see also \cite{Dormy2018}): he suggests that an efficient strategy to achieve strong-field dynamo saturation in DNS is to reduce the Ekman number even more rapidly than the magnetic Prandtl number $\Pm=\nu/\eta$ (see the discussion section). 
In the following we assume that these criteria are met, and we neglect the viscous contributions to the nonlinear $\alpha$-effect. Removing the terms involving the Reynolds number leads to the simpler form:
\begin{eqnarray}
{\alpha}_{xx;yy} =  \frac{-\Rm^2}{ \frac{{\cal B}_{x;y}^4 }{R^2 (Ro^{-1}-1)^2} +2 \frac{{\cal B}_{x;y}^2}{Ro^{-1}-1} +1+R^2}      \, ,\label{alphaVsimple}
\end{eqnarray}
We now study the magnetic field arising through this nonlinear $\alpha$-effect.

\section{Linear instability and vicinity of the dynamo threshold \label{seclinstabWNL}}

The $x$ and $y$ components of the large-scale induction equation (\ref{lsind}) are:

\begin{eqnarray}
\partial_{T} \Bx & = & - \epsilon^{-1} \, \partial_{Z}  ({\alpha}_{yy} \By) + \partial_{ZZ} \Bx \, , \\
\partial_{T} \By & = &  \epsilon^{-1}  \, \partial_{Z}  ({\alpha}_{xx} \Bx) + \partial_{ZZ} \By \, , 
\end{eqnarray}
where the inviscid $\alpha$-effect coefficients are given by (\ref{alphaVsimple}). The fields and coefficients appearing in these two equations depend on $Z$ and $T$ only. To alleviate the algebra, we therefore unambiguously switch back to the standard unscaled variables $z$ and $t$, using $\partial_{T}=\epsilon^{-2} \, \partial_{t} $ and $\partial_{Z}=\epsilon^{-1} \, \partial_{z}$. In terms of these unscaled variables, the fields $\Bx(z,t)$ and $\By(z,t)$ obey the following set of equations:
\begin{eqnarray}
\partial_{t} \Bx & = & - \partial_{z}  ({\alpha}_{yy} \By) + \partial_{zz} \Bx \, , \label{eqBx}\\
\partial_{t} \By & = &    \partial_{z}  ({\alpha}_{xx} \Bx) + \partial_{zz} \By \, . \label{eqBy}
\end{eqnarray}
The remainder of the analysis is concerned with the solutions to this set of equations, with particular emphasis on the scaling behaviour of the magnetic energy above the dynamo threshold.

\subsection{linear instability \label{seclinstab}}

Let us first study the linear stability of the system of equations (\ref{eqBx}-\ref{eqBy}): we consider infinitesimal perturbations $(\Bx,\By)\ll 1$ and linearize the equations by substituting the expression (\ref{alphaVsimple}) of the  $\alpha$-effect coefficients evaluated for $\Bx=\By=0$. This leads to a linear set of equations with $z$-independent coefficients:
\begin{eqnarray}
\partial_{t} \Bx & = & \frac{\Rm^2}{1+R^2} \, \partial_{z}  \By + \partial_{zz} \Bx \, ,\\
\partial_{t} \By & = & - \frac{\Rm^2}{1+R^2} \, \partial_{z}   \Bx + \partial_{zz} \By \, , 
\end{eqnarray}
 the solution to which can be sought in the form of a single Fourier mode in $z$. We therefore introduce the complex variable:
\begin{eqnarray}
\Bx +i \By = A(t) \exp\left( i \frac{2 \pi \ell}{\lambda} z \right) \, ,
\end{eqnarray}
where $\lambda/\ell$ is the dimensionless vertical wavelength of the perturbation, $\lambda$ being the dimensional one. The dynamo threshold is attained when the set of linear equations admits nonzero time-independent solutions for $A$, which leads to the following critical magnetic Reynolds number $\Rm_c$ for linear instability:
\begin{eqnarray}
\Rm_c \sqrt{\frac{\lambda}{\ell}} = \sqrt{2\pi (1+R^2)} \, .\label{eqRmc}
\end{eqnarray}
As in the standard G.O. Roberts dynamo, the threshold for instability is best expressed in terms of a critical magnetic Reynolds number based on the harmonic mean $\sqrt{\lambda \ell}$ between the small and large scales: $\Rm \sqrt{\lambda/\ell}=V^* \sqrt{\lambda \ell}/\eta$.  The large-scale sweeping flow is detrimental to the linear instability: the threshold (\ref{eqRmc}) for linear instability increases with the magnetic Reynolds number $R$ associated to the large-scale velocity ${\cal U}$. It is interesting to compare this result to those of \citet{Tilgner}, who studied the same kinematic dynamo problem but focused on small-scale modes with $\lambda \sim \ell$: for such small-scale dynamo action, strong sweeping is also detrimental to the dynamo effect, but a weak sweeping flow was shown to decrease the threshold magnetic Reynolds number.

\subsection{Saturation near the dynamo threshold: strong-field scaling regime}

In the vicinity of the dynamo threshold, the saturation of the dynamo instability is governed by the quadratic term in ${\cal B}$ at the denominator of (\ref{alphaV}). If this term is positive, the magnitude of the $\alpha$-effect is reduced as the field grows, leading to a supercritical bifurcation. By contrast, if this term is negative, the first nonlinearities do not saturate the instability and we expect a subcritical bifurcation. These arguments can be made more precise by computing the normal form in the vicinity of the instability threshold, using  standard asymptotic methods. The multiple-scale expansion is described in appendix \ref{appnormalform} and leads to:
\begin{eqnarray}
\frac{\mathrm{d}{A}}{\mathrm{d}t} = \frac{4 \sqrt{2} \pi^{3/2}}{\sqrt{1+R^2}} \left( \frac{\ell}{\lambda}\right)^{3/2} (\Rm-\Rm_c) \, A - \frac{6 \pi^2}{(1+R^2)(Ro^{-1}-1)} \left( \frac{\ell}{\lambda}\right)^{2} A^2 \bar{A} \, . \label{normalform}
\end{eqnarray}

The nature of the bifurcation crucially depends on the sign of $Ro^{-1}-1$: 
\begin{itemize}
\item For $Ro^{-1}<1$, the cubic term in (\ref{normalform}) does not saturate the instability and the bifurcation is subcritical. One way to study the dynamo saturation would be to push this expansion to higher order, hoping that the next nonlinear term saturates the instability. However, for some parameter values even the fifth-degree monomial in $A$ does not saturate the instability. In the next section, we therefore follow another route than perturbative expansion and directly compute the expression of the steady dynamo branch, which remains valid at finite distance from threshold.
\item For $Ro^{-1}>1$, the cubic nonlinearity saturates the instability, which therefore becomes a supercritical pitchfork bifurcation.
\end{itemize}
Seeking stationary solutions to the normal form in the latter case, we obtain the magnetic energy in the vicinity of the instability threshold:
\begin{eqnarray}
|A|^2 = \frac{2}{3} \sqrt{\frac{2(1+R^2)}{\pi}}    \, |Ro^{-1}-1|  \, \sqrt{\frac{\lambda}{\ell}} \, (\Rm-\Rm_c)  \, . 
\end{eqnarray}
In the large rotation limit $|Ro|\ll 1$, this corresponds to the strong-field scaling regime, where the magnetic energy is proportional to $\omega$ and independent of $\nu$:
\begin{eqnarray}
\frac{B^{*2} \ell^2}{\rho \mu_0 \eta^2} \sim  \frac{\ell \omega}{{\cal U}} \sqrt{1+R^2} \, \sqrt{\frac{\lambda}{\ell}} \, (\Rm-\Rm_c) \, .
\end{eqnarray}
We stress the fact that in these relations $\eta$ can be replaced by $V^* \ell \sqrt{\lambda/\ell}$, using the fact that $\Rm \simeq \Rm_c$.
For instance, if $R \gg 1$, we can rewrite this expression in the simpler form:
\begin{eqnarray}
\frac{B^{*2} }{\rho \mu_0 } \sim  V^* \, \lambda \, \omega \, (\Rm-\Rm_c) \, . \label{llsf}
\end{eqnarray}
\cor{This expression corresponds to the scaling-law proposed by \citet{Petrelis2001} for supercritical dynamos saturating through the action of the Coriolis force. As noted by these authors, the scaling-law (\ref{llsf}) resembles the strong-field scaling regime in the sense that the magnetic energy is proportional to the global rotation rate $\omega$ and independent of viscosity. However, because the scaling-law is valid close to threshold, a key difference between (\ref{llsf}) and a strong-field dynamo regime is that the Lorentz force is much weaker than both the Coriolis term and the body-force driving the fluid (because of the factor $Rm-Rm_c$ in (\ref{llsf})). By contrast, subcritical bifurcations do arise in this model when $Ro^{-1}<1$, and when $Ro^{-1}$ has a large negative value the resulting dynamo branch is in Magnetic-Forcing-Coriolis balance, as expected for a body-forced strong-field dynamo branch. In the next section we compute the bifurcated dynamo branches at arbitrary distance from threshold, therefore shedding light on the saturation of these subcritical dynamos. We will see that both the supercritical and subcritical dynamos achieve MFC balance at large distance from threshold, provided $|Ro| \ll 1$.}

%

\section{Dynamo branches at finite distance from threshold \label{secfinite}}

In contrast with most existing analytical nonlinear dynamo models, our approach does not require the magnetic Reynolds number to be close to threshold. This opens an analytical avenue to study dynamo saturation at large distance from threshold, the goal being twofold: first, we will characterise the subcritical dynamo branches identified in the previous section. Second, we will study the behavior of the magnetic energy far away from the dynamo threshold, providing an analytical example of equipartition between kinetic and magnetic energy \cor{in the absence of global rotation, and an example of the magneto-geostrophic scaling regime of \citet{PHRoberts72} for rapid rotation.}



As a word of caution, we stress the fact that we focus on the saturation of the large-scale dynamo. The precise regime in which the computation is valid is $\Rm \ll 1$, $\Rm \sqrt{\lambda/\ell} \sim 1$. The flow is then subject to a large-scale dynamo instability only, that arises through the $\alpha$-effect (\ref{alphaVsimple}). In particular, this range of parameters rules out small-scale dynamo action: the latter arises for even faster flows with $\Rm=\mathcal{O}(1)$ and is characterised by the growth and saturation of magnetic field modes at the scale $\ell$ of the cellular flow \citep{Vainshtein,Cattaneo,Tilgner,Ponty,Seshasayanan2016,Cameron}.

\begin{figure}
    \centerline{\includegraphics[width=7 cm]{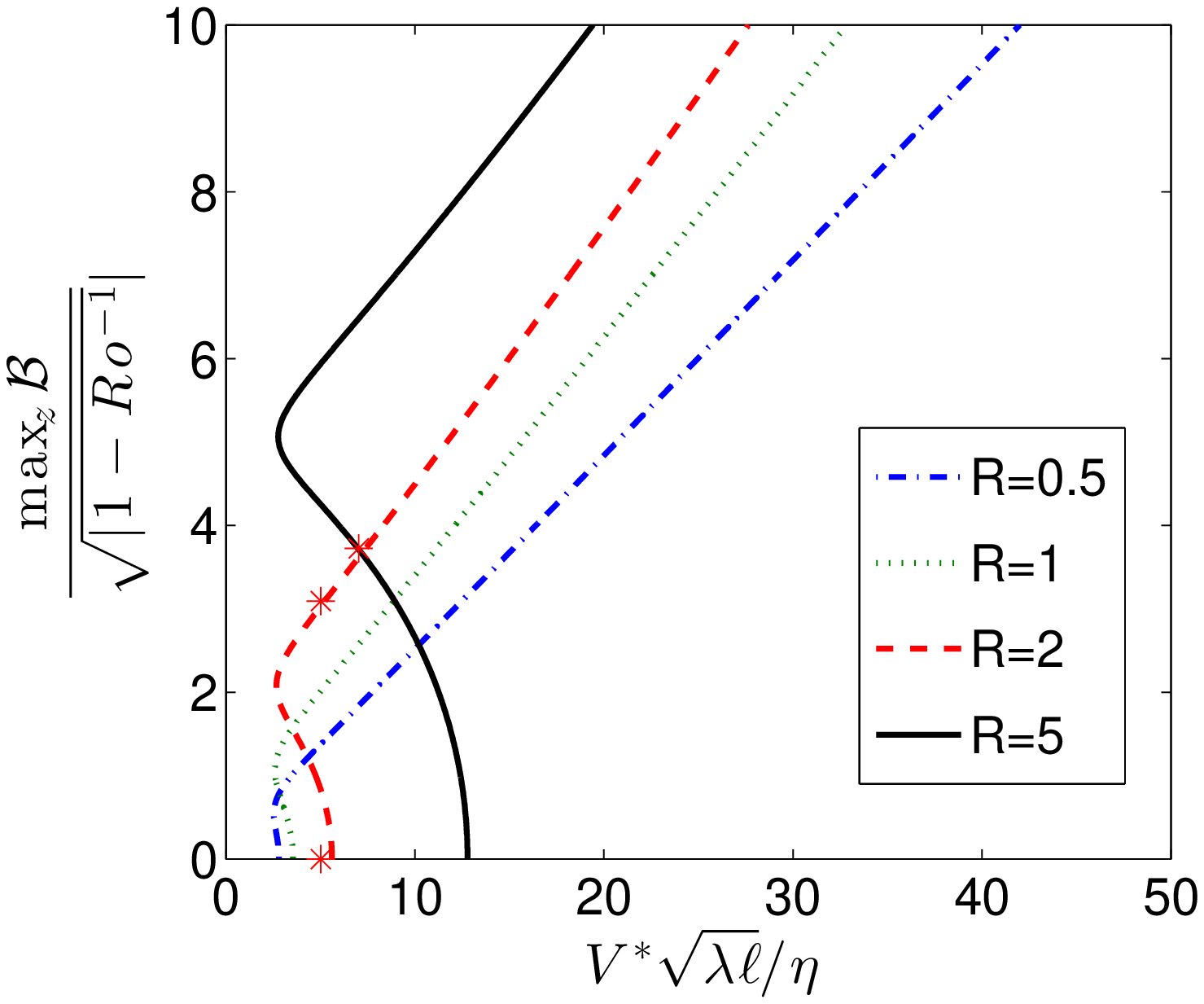} \includegraphics[width=7 cm]{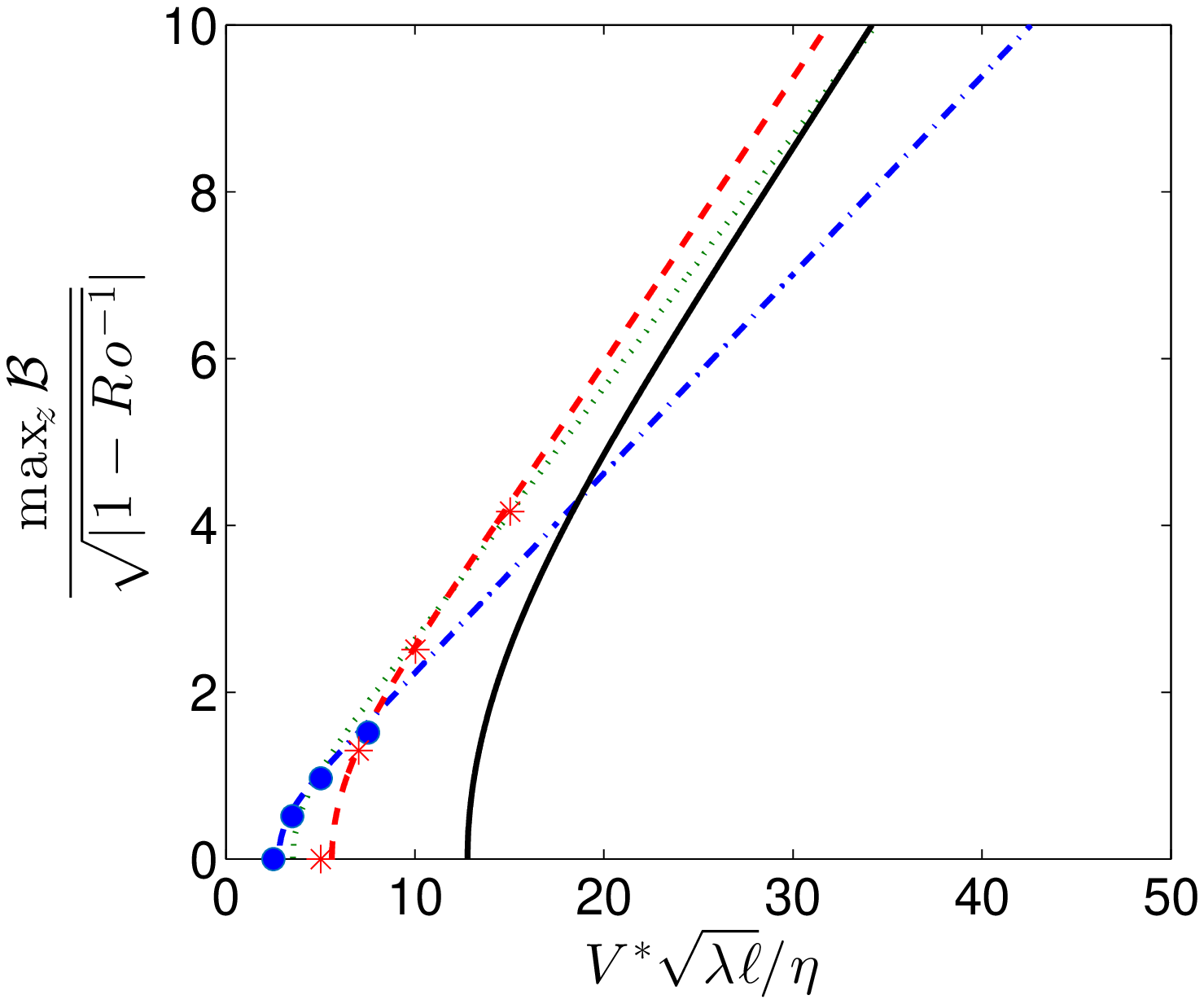}}
      \caption{Dynamo bifurcation curves at fixed $R$. \textbf{Left-hand panel:} $Ro^{-1}<1$. The dynamo bifurcation is subcritical. For $\Rm$ such that two nonzero values of ${\cal B}$ are solution, the lower value is unstable while the greater one is stable. \textbf{Right-hand panel:} $Ro^{-1}>1$. The dynamo bifurcation is supercritical. Symbols are results from numerical simulations: $\bullet$, $R = 0.5$; $*$, $R = 2.0$ (see appendix \ref{app:Numsimuls} for details).\label{biffixedR}}
\end{figure}

\begin{figure}
    \centerline{\includegraphics[width=14 cm]{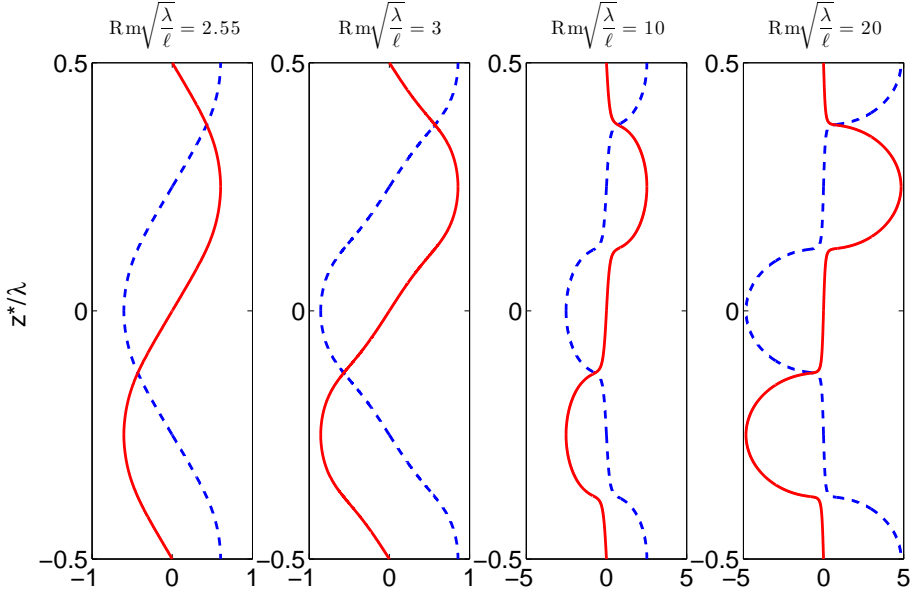}}
      \caption{Saturated dynamo state for $R=0.5$ and several values of $\Rm$, in the case  $Ro^{-1}<1$. solid line, ${\cal B}_x/\sqrt{1-Ro^{-1}}$; dashed line, ${\cal B}_y/\sqrt{1-Ro^{-1}}$. \label{mode}} 
\end{figure}

\subsection{Bifurcated dynamo branches}

We look for steady solutions to equations (\ref{eqBx}-\ref{eqBy}). After one integration in $z$ we obtain:
\begin{eqnarray}
\frac{d \Bx}{dz} & = &  \frac{- \Rm^2 R^2 \By}{\frac{\By^4}{(Ro^{-1}-1)^2} +2 R^2 \frac{\By^2}{Ro^{-1}-1} +R^2(1+R^2)} \label{eqstatBx}\\
\frac{d \By}{dz} & = &  \frac{ \Rm^2 R^2 \Bx}{\frac{\Bx^4}{(Ro^{-1}-1)^2} +2 R^2 \frac{\Bx^2}{Ro^{-1}-1} +R^2(1+R^2)} \, . \label{eqstatBy}
\end{eqnarray}
The integration constants have been set to zero. This is a necessary condition if one integrates the equations over one spatial period in $z$, demanding that the eigenmode transforms as ${\cB} \to -{\cB}$ when shifted by half a period in $z$.

We first focus on the case $Ro^{-1}<1$, the changes to be made when $Ro^{-1}>1$ being discussed after equation (\ref{eqbranch}). Dividing the two equations by $\sqrt{1-Ro^{-1}}$ makes it clear that the magnetic field and Rossby number only enter the equations through the combinations $G_{x;y}={\cal B}_{x;y}/\sqrt{1-Ro^{-1}}$, which already shows that the saturated magnetic energy will depend on the Rossby number only through a prefactor $1-Ro^{-1}$. $G_x(z)$ and $G_y(z)$ obey the system of equations:
\begin{eqnarray}
\frac{d G_x}{dz} & = &  \frac{- \Rm^2 R^2 G_y}{G_y^4 -2 R^2 G_y^2 +R^2(1+R^2)} \label{eqstatGx}\\
\frac{d G_y}{dz} & = &   \frac{ \Rm^2 R^2 G_x}{G_x^4 -2 R^2 G_x^2 +R^2(1+R^2)} \, . \label{eqstatGy}
\end{eqnarray}

Upon multiplying (\ref{eqstatGx}) and (\ref{eqstatGy}) we obtain:
\begin{equation}
 \frac{  G_x}{G_x^4 -2 R^2 G_x^2 +R^2(1+R^2)} \frac{d G_x}{dz}  - \frac{ G_y}{G_y^4 -2 R^2 G_y^2 +R^2(1+R^2)}  \frac{d G_y}{dz}  = 0 \, ,
\end{equation}
which we integrate into:
\begin{equation}
\arctan \left( \frac{G_x^2}{R} -R  \right) + \arctan \left( \frac{G_y^2}{R} -R  \right)= \text{const.} \, ,
\end{equation}
and after taking the tangent, using the formula for $\tan(a+b)$ and rearranging leads to
\begin{equation}
(G_x^2+G_y^2) \left( \frac{1}{R} - {\cal C} \right) = 2 R + {\cal C} \left(1-R^2- \frac{G_x^2 G_y^2}{R^2} \right) \, ,\label{1stintegral}
\end{equation}
where ${\cal C}$ is a $z$-independent constant. To determine its value, we denote as $M$ the maximum magnitude attained by $G_x$ (and $G_y$) over one oscillation in $z$. Because $G_x$ and $G_y$ are in quadrature, $G_y$ vanishes when $G_x=M$. Substituting into (\ref{1stintegral}) we obtain ${\cal C}$ as a function of $M$:
\begin{equation}
{\cal C} = \frac{M^2-2R^2}{R(1+M^2)-R^3} \, . \label{valC}
\end{equation}
From equation (\ref{1stintegral}) we extract $G_x$ as a function of $G_y$:
\begin{equation}
G_x = \pm \sqrt{\frac{G_y^2 ({\cal C}-1/R)+2R+{\cal C}(1-R^2)}{1/R-{\cal C} +{\cal C} G_y^2 /R^2}} \, .
\end{equation}
Substituting this expression into the right-hand side of (\ref{eqstatGy}) leads to a differential equation where the variables $z$ and $G_y$ can be separated. We can then integrate this expression to get $z$ as a function of $G_y$, with the boundary condition $G_y(z=0)=0$. The resulting expression gives the spatial structure of the dynamo magnetic field.

If we set $z=\lambda/4 \ell$, where $\lambda$ still denotes the dimensional wavelength along $z$, then $G_y=M$. We therefore obtain $\Rm^2 \lambda/\ell$ as a function of $M=\max_z\{\Bx\}/\sqrt{1-Ro^{-1}}$, which is the bifurcation curve we are looking for:

\begin{eqnarray}
\nonumber \Rm^2 \, \frac{\lambda}{\ell} & = & \frac{8 i \sqrt{R^4+R^2} (M^4+R^2-2M^2 R^2+R^4) M^2}{R^2(M^4-2M^2R^2)^{3/2}} \\
\nonumber & & \times \left[ M^2 {\cal E}\left(i\sqrt{\frac{M^4-2M^2R^2}{R^2+R^4}} \, ; i \sqrt{\frac{R^2+R^4}{M^4-2M^2R^2}} \right) \right.  \\
 & & + \left. (R^2-M^2) {\cal F}\left(i\sqrt{\frac{M^4-2M^2R^2}{R^2+R^4}} \, ; i \sqrt{\frac{R^2+R^4}{M^4-2M^2R^2}} \right) \right] \, ,
 \label{eqbranch}
\end{eqnarray}
where ${\cal F}$ and ${\cal E}$ are the incomplete elliptic integrals of the first and second kinds in Jacobi's form, whose precise definitions are given in appendix \ref{appelliptic}. Expression (\ref{eqbranch}) above is valid when $Ro^{-1}<1$. The expression for $Ro^{-1}>1$ is obtained by substituting $M^2=-N^2$ in expression (\ref{eqbranch}), with $N=\max_z\{\Bx\}/\sqrt{|Ro^{-1}-1|}$.

The square root of (\ref{eqbranch}) is the reciprocal of the bifurcation curve. From this expression, we can plot the bifurcation curves $\max_z\{\Bx\}$ vs $\Rm \sqrt{\lambda / \ell}$. Examples of such curves are shown in figure \ref{biffixedR} for both signs of $Ro^{-1}-1$. As expected, the dynamo is subcritical for $Ro^{-1}<1$ and supercritical for $Ro^{-1}>1$. In both cases the departure from $M=0$ is well captured by the normal form (\ref{normalform}). The magnetic field structure is displayed in figure \ref{mode}: close to onset, both components are sinusoidal in $z$, in agreement with the analysis in section \ref{seclinstab}. As we move further away from onset the magnetic field becomes more and more anharmonic as a consequence of the nonlinearities. To confirm the theoretical results, we have performed a few direct numerical simulations of the complete MHD equations \ref{dimlessNS}-\ref{dimlessInd}. The details of the numerical code and parameters used are given in Appendix \ref{app:Numsimuls}. After some transient, these simulations reach a steady state. The symbols in figure \ref{biffixedR} indicate the magnitude of the corresponding magnetic field.


\subsection{Scaling behavior of the magnetic energy}

\begin{figure}
    \centerline{\includegraphics[width=9 cm]{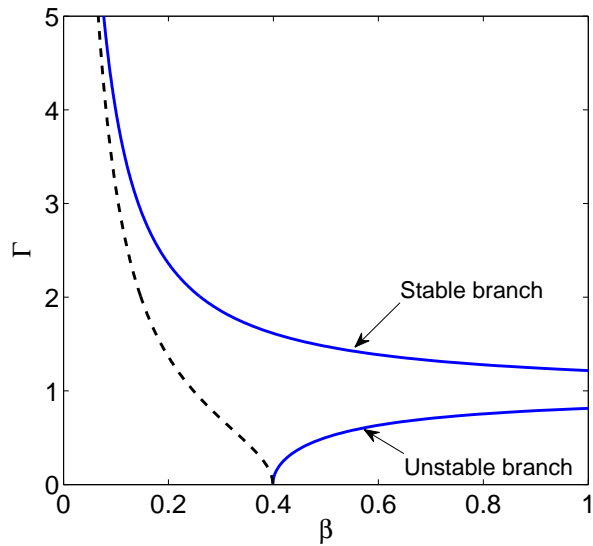}}
      \caption{Asymptotic limit of low resisitivity: prefactor $\Gamma$ of the scaling-law (\ref{scalingloweta}) for the magnetic energy, as a function of the velocity ratio $\beta={\cal U}/(V^* \sqrt{\lambda/\ell})$. For $Ro^{-1}>1$ (dashed line), the flow induces a dynamo for $\beta<1/\sqrt{2\pi}$ only. For $Ro^{-1}<1$ (solid line), the flow induces a dynamo for $\beta<1/\sqrt{2\pi}$, while it is bistable between a dynamo and a non-dynamo state for $\beta>1/\sqrt{2\pi}$. \label{figloweta}} 
\end{figure}

Close to the threshold of a supercritical dynamo bifurcation, the magnetic energy crucially depends on the magnetic diffusivity: a slight change in magnetic diffusivity has a strong impact on the distance from the dynamo threshold, and therefore on the magnetic energy. By contrast, when the dynamo bifurcation is subcritical, or when the system is far away from threshold, the situation is less clearly established: does the magnetic energy still depend strongly on the magnetic diffusivity? Or does it reach a regime where magnetic diffusivity is irrelevant, in a similar fashion to kinematic viscosity in standard hydrodynamic turbulence? 

Consider the subcritical dynamo branches in the left panel of figure (\ref{biffixedR}). The value of $\max_z\{\Bx\}/\sqrt{|Ro^{-1}-1|}$ at the beginning of the dynamo branch (the leftmost point of each curve) scales with $R$. For rapid global rotation and in terms of dimensional quantities, we obtain:
\begin{equation}
\frac{B^{*\, 2}}{\rho \mu_0 \eta \omega} \sim R \, .
\end{equation}
For $R={\cal O}(1)$, this corresponds to an Elsasser number of the order of unity. However, for arbitrary $R$, substituting the definition of $R$ leads to:
\begin{equation}
\frac{B^{*\, 2}}{\rho \mu_0 \ell {\cal U}  \omega} \sim 1 \, . \label{modifiedElsasser}
\end{equation}
This dimensionless number is a ``turbulent'' Elsasser number in which the magnetic diffusivity has been replaced by an effective diffusivity $\ell {\cal U}$ based on the sweeping velocity. For rapid global rotation, this Elsasser number is of the order of unity on the subcritical dynamo branch. The relation (\ref{modifiedElsasser}) corresponds to the ratio of magnetic to kinetic energy being given by the inverse Rossby number, $Ro^{-1}$. \cor{This scaling-law is called ``magneto-geostrophic'' in \citet{PHRoberts72}. Coming back to the Navier-Stokes equation (\ref{dimlessNS}) and its solution (\ref{expru}-\ref{exprwy}), one can check that for $Ro\ll 1$ the dominant balance is then between the Coriolis term, body-force and Lorentz force: this is the Magnetic-Forcing-Coriolis balance.}

In the absence of global rotation, $Ro^{-1}=0$, the scaling relation (\ref{modifiedElsasser}) for the subcritical dynamo branch is replaced by:
\begin{equation}
\frac{B^{*\, 2}}{\rho \mu_0 \, {\cal U}^2} \sim 1 \, , \label{eqreg}
\end{equation}
which is the regime of equipartition between magnetic energy and kinetic energy.

To put these scaling laws on firm analytical ground, we focus on the asymptotic behavior of the dynamo branches at large distance from threshold. Indeed, our asymptotic model allows us to reach a regime where the conductivity is large enough for the large-scale dynamo to be far away from threshold, but small enough to prevent any small-scale dynamo action: $\Rm \ll 1$, but $\Rm  \sqrt{\lambda/\ell} \gg 1$. In this regime, we wish to show that the magnetic energy behaves as:
\begin{eqnarray}
\max_z\{\Bx\}/\sqrt{|Ro^{-1}-1|} \simeq \Gamma R \, , \label{scalingloweta}
\end{eqnarray}
where $\Gamma$ is a constant. We denote as $\beta$ the following ratio of the magnetic Reynolds numbers:
\begin{eqnarray}
\beta = \frac{R}{\Rm \sqrt{\lambda/\ell}} = \frac{{\cal U}}{V^* \sqrt{\lambda/\ell}}
\end{eqnarray}
The limit of large distance from threshold is taken by considering $|M| \gg 1$, $R \gg 1$ and $\Rm  \sqrt{\lambda/\ell} \gg 1$, keeping the ratios $\Gamma$ and $\beta$ constant. We stress the fact that this regime can only be achieved for very small values of $\epsilon$, in order to maintain the asymptotic ordering: for instance, quantities that are ${\cal O}(1)$ in the expansion can be large, as long as they remain much smaller than $\epsilon^{-1/2}$. In this limit and for $Ro^{-1}<1$, equation (\ref{eqbranch}) gives:
\begin{eqnarray}
\nonumber \beta = \left| \frac{(\Gamma^2-2)^{\frac{3}{4}}}{2 \sqrt{2 \Gamma} |\Gamma^2-1| } \left[ i {\cal E}\left(i \Gamma \sqrt{\Gamma^2-2} \, ; \frac{i}{\Gamma \sqrt{\Gamma^2-2}} \right) + \frac{i (1-\Gamma^2)}{\Gamma^2} {\cal F}\left(i \Gamma \sqrt{\Gamma^2-2} \, ; \frac{i}{\Gamma \sqrt{\Gamma^2-2}}  \right)  \right]^{-\frac{1}{2}}  \right| .\\
\label{eqbeta}
\end{eqnarray}
The corresponding expression for $\beta$ in the case $Ro^{-1}>1$ is obtained by substituting $\Gamma \to i \Gamma$ in expression (\ref{eqbeta}).

For a given value of the velocity ratio $\beta$, the relation above can be inverted to extract the prefactor $\Gamma$ of the scaling-law (\ref{scalingloweta}) for the magnetic energy. This proves that the approach is sound and confirms the ansatz (\ref{scalingloweta}). In figure \ref{figloweta} we plot the prefactor $\Gamma$ as a function of the velocity ratio $\beta$ for the two signs of $Ro^{-1}-1$. For $Ro^{-1}>1$, the prefactor $\Gamma$ differs from zero only for $\beta < 1/\sqrt{2 \pi}$. This is because for $\beta > 1/\sqrt{2 \pi}$ the system remains stable to the dynamo instability regardless of the value of $\eta$, see expression (\ref{eqRmc}) for the dynamo threshold. The situation for $Ro^{-1}<1$ is different: for $\beta > 1/\sqrt{2 \pi}$, the system is linearly stable to magnetic perturbations, but a stable subcritical dynamo branch coexists with the non-dynamo branch ${\bf B}={\bf 0}$. The basins of attraction of these two stable states are separated by an unstable dynamo branch, see figure \ref{figloweta}. This study therefore highlights the crucial role of large-scale sweeping flows in hindering the dynamo effect: for strong enough sweeping the system becomes linearly stable to magnetic perturbations, although subcritical dynamo states exist for $Ro^{-1}<1$.

More than the precise value of this prefactor, it is the scaling behavior of the magnetic energy that is of interest to us. We obtain:
\begin{eqnarray}
\frac{B^*}{\sqrt{\rho \mu_0}} \sim {\cal U} \sqrt{|Ro^{-1}-1|} \, , \label{scalingsmalleta}
\end{eqnarray}
which shows clearly that the magnetic energy is independent of magnetic diffusivity. \cor{In the case where the fluid is not rotating this relation reduces to the equipartition scaling regime (\ref{eqreg}), while in the limit of rapid global rotation it reduces to the magneto-geostrophic scaling relation (\ref{modifiedElsasser}), characterized by Magnetic-Forcing-Coriolis balance.}

%


\section{Discussion \label{secdisc}}

\cor{We have introduced simple dynamo flows exhibiting the ``strong-field'' scaling-law for the saturated magnetic energy. Using a combination of global rotation, large-scale sweeping flow and small-scale forcing, we showed that the magnetic energy is independent of viscosity when the latter is small enough, and proportional to the rotation rate for rapid rotation. Of course, because the flow is driven by a body-force and not by thermal convection, we do not reproduce the multiple-branch picture conjectured by Roberts \citep{Roberts78,Roberts88}. In particular, the MAC balance of a convective strong-field dynamo is replaced here by a Magnetic-Forcing-Coriolis balance, which yields the magneto-geostrophic scaling-law for the magnetic energy. }Depending on the relative directions of global rotation and large-scale sweeping flow, the dynamo transition is either subcritical or supercritical. We are not aware of other analytical examples of subcritical dynamos: here the large-scale sweeping flow seems to be the key ingredient for subcriticality.

As opposed to standard weakly nonlinear methods \citep{Nunez,Seshasayanan}, our study is based on scale separation only and is not restricted to the immediate vicinity of the dynamo threshold. We therefore studied the scaling behavior of the magnetic energy at large distance from threshold: when both the Reynolds number and magnetic Reynolds number are large, the magnetic energy is independent of both viscosity and magnetic diffusivity. In the absence of global rotation, the resulting scaling-law corresponds to equipartition between kinetic and magnetic energy. With global rotation, the ratio of kinetic to magnetic energy -- the squared Alfvén number -- is proportional to the Rossby number. \cor{This corresponds again to the magneto-geostrophic scaling-law \citep{PHRoberts72}, with a ``turbulent'' Elsasser number of the order of unity.}

\cor{The following subsections discuss the case of a large-scale zonal flow, the stability of the analytical dynamo branches, and the criteria to achieve the strong-field regime in DNS.}

\subsection{The case of a sweeping zonal flow \label{secsweeping}}

The main body of the present study deals with the situation where the global rotation $\bOmega$ and the large-scale sweeping flow $\bcU$ are collinear. We focused on this situation because it leads to a variety of bifurcations, the dynamo being either subcritical or supercritical. However, a situation of important astrophysical relevance is that of a large-scale zonal flow, perpendicular to the global rotation vector $\bOmega$ \citep{Aubert2005,Gomez,Schrinner}. We therefore reproduced the present computations for a situation similar to that of figure \ref{schema}, except that the large-scale flow is now perpendicular to $\bOmega$: we write $\bcU = {\cal U}  (\textbf{e}_x - \textbf{e}_y)$. The analysis is similar to the case developed above and we only state the main results. 

The dimensionless $\alpha$-effect coefficients are: 
\begin{eqnarray}
{\alpha}_{xx} =  - \Rm^2 \, \frac{(Ro^2-1)^2}{Ro^2+1} \times  \frac{1}{ (1+R^2)(1-Ro)^2+2 \Bx^2 Ro(1-Ro)+ \Bx^4 Ro^2  /R^2  }      \, ,  \\
{\alpha}_{yy} =    - \Rm^2 \, \frac{(Ro^2-1)^2}{Ro^2+1} \times  \frac{1}{ (1+R^2)(1+Ro)^2-2 \By^2 Ro(1+Ro)+ \By^4 Ro^2  /R^2  }     \, .
\end{eqnarray}
Denoting the critical magnetic Reynolds number as $\Rm_c^{(\text{zonal})}$, we obtain through linear stability analysis: 
\begin{eqnarray}
\Rm_c^{(\text{zonal})} \sqrt{\frac{\lambda}{\ell}} = \sqrt{2 \pi (1+R^2)} \times \sqrt{\frac{Ro^2+1}{|Ro^2-1|}} \, .
\end{eqnarray}
The base-flow differs from the standard G.O. Roberts flow and $\Rm_c^{(\text{zonal})}$ now explicitly depends on the Rossby number. Once again, we can determine the nature of the dynamo bifurcation using standard weakly nonlinear analysis. A straightforward computation of the normal form shows that the dynamo bifurcation is always subcritical in the presence of a large-scale zonal flow.

\subsection{Stability properties and turbulent regime}

When analysis is pushed into the low-viscosity regime, a fair question arises as to whether the corresponding flows are stable. In the present situation, we stress the fact that there is indeed a region of parameter space where our dynamo solutions should be stable. First of all, the present flow is not subject to the kinetic-alpha-effect \citep{Frisch}, and its stability properties are therefore independent of the scale separation $\lambda / \ell$. Instead, the flow goes unstable through a negative-viscosity mechanism  when the small-scale Reynolds number exceeds a threshold of the order of unity \citep{Sivashinsky}. Provided $V^* \ell / \nu \lesssim 1$, the hydrodynamic flow should therefore remain linearly stable. When the scale separation $\lambda/\ell$ is large enough, this viscous small-scale flow can trigger the dynamo instability discussed above, the resulting magnetic energy being independent of viscosity at low Ekman number $\nu / \ell^2 \omega$  and/or large sweeping Reynolds number $Re={\cal U} \ell / \nu$.

Another source of deviations from the computed dynamo branches could be secondary instabilities from the bifurcated solution. Such instabilities probably arise at large distance from threshold. However, we checked using a standard pseudo-spectral solver that the solution to the full MHD equations (\ref{dimlessNS}-\ref{dimlessInd}) indeed corresponds to the branch we computed at moderate distance from threshold. \cor{The full domain of stability of our solutions could be investigated through extensive DNS, or possibly analytically, using the approach of \citet{Courvoisier}.}
Even if the strong-field dynamo branches of the present study did become unstable in some region of parameter space, it is very unlikely that viscosity would come back into play, and the magnetic energy should keep displaying a clear strong-field scaling regime.
 

\subsection{Criteria to achieve the strong-field regime: low $Pm$ versus low Ekman number }

There is currently a debate over the optimal strategy to reach astrophysically relevant regimes in dynamo DNS. While the natural approach would be to try to reach low magnetic Prandtl numbers, \citet{Dormy} suggested that the Ekman number should be lowered even more rapidly than $\Pm$. It is interesting to notice that the criterion to achieve the strong-field regime in the present study is precisely that of a low-Ekman number (see section \ref{secneglectviscous}). Whether Ekman or $\Pm$ is the right parameter in fact depends very much on the geometry of the forcing: in a previous study \citep{Seshasayanan}, we considered a forcing that is compatible with the Taylor-Proudman constraint -- i.e., a forcing that is invariant along $\bOmega$ \citep{Gallet2015}-- and showed that the right criterion to neglect viscosity is low magnetic Prandtl number, $\Pm \ll 1$. The magnetic field then achieves the inertial or turbulent scaling regime: the magnetic energy is independent of viscosity but also of the global rotation rate. By contrast, in the present situation the forcing directly shears $\bOmega$ and is therefore incompatible with the Taylor-Proudman constraint \citep{Campagne}. The right criterion to neglect viscosity becomes low Ekman number, and the magnetic energy obeys the ``strong-field'' scaling regime. 

\cor{We can summarize the findings of \citet{Seshasayanan} and of the present study as follows: geostrophic base flows lead to the inertial scaling regime, with $B^2$ independent of the rotation rate, whereas base flows that are not in geostrophic balance can achieve the magneto-geostrophic scaling regime, with much larger magnetic energy, proportional to the global rotation rate. In spherical geodynamo simulations, which criterion should be retained to observe a $\nu$-independent scaling regime -- and whether this scaling-law involves the global rotation rate -- may depend on the region of the sphere that contributes most to magnetic-field generation.} 
Low-$\Pm$ might be needed wherever the flow is quasi-2D (typically outside the tangent cylinder), whereas low-Ekman-number may be the right criterion wherever the flow varies rapidly along the axis of rotation (inside the tangent cylinder, see \citet{Schaeffer}). The strong-field scaling regime would then arise from dynamo saturation inside the tangent cylinder.

\vspace{0.5 cm}

This research is supported by the European Research Council under grant agreement FLAVE 757239, and by ANR ``Excellence laboratory'' grant ANR-10-LABX-0039.




\appendix

\section{Derivation of the reduced equations \label{appeq}}

Expand the magnetic field as:
\begin{eqnarray}
{\bf B} = {\hbB}_0 (x,y,Z,t,T) + \epsilon^{1/2} {\hbB}_{1/2} (x,y,Z,t,T) + \epsilon {\hbB}_{1} (x,y,Z,t,T) + \dots \, ,
\end{eqnarray}
where the quantities with a hat are $\mathcal{O}(1)$ and independent of $\epsilon$. The velocity field is scaled as:
\begin{eqnarray}
{\bf v} =  \epsilon^{1/2} {\hbv}_{1/2} (x,y,Z,t,T) + \epsilon {\hbv}_{1} (x,y,Z,t,T) + \dots \, .
\end{eqnarray}
and the forcing amplitude as $F= \epsilon^{1/2} \hat{F}$. The parameters $R$ and $Ro$ are $\mathcal{O}(1)$. The time derivative and gradient operators become:
\begin{eqnarray}
\partial_t & = & \partial_t + \epsilon^2 \partial_T \, , \\
\bnabla & = & \bnabla_{\bf x} + \epsilon \bnabla_{\bf X} \, . 
\end{eqnarray}
because we are using fast horizontal variables scales and a slow vertical one, the Laplacian operator simplifies to:
\begin{eqnarray}
\bnabla^2 = \bnabla_{\bf x}^2 + \epsilon^2 \bnabla_{\bf X}^2 \, . 
\end{eqnarray}
Collecting the terms of order ${\cal O}(1)$, the induction equation (\ref{dimlessInd}) yields:
\begin{eqnarray}
\partial_t {\hbB}_0 + R [({\bf e}_x + {\bf e}_y) \cdot \bnabla_{\bf x}] {\hbB_0} &  = & \bnabla_{\bf x}^2 {\hbB_0} \, . \label{order0}
\end{eqnarray}
This is an unforced advection diffusion equation for $\hbB_0$. After a transient on the short timescale $t$, $\hbB_0$ becomes independent of the small-scale variables $x$ and $y$, and therefore of $t$. Hence we write the solution in the long time $t$ limit as:
\begin{eqnarray}
{\hbB}_0 = \cB(Z,T) \, .
\end{eqnarray}
Collecting the terms of order $\epsilon^{1/2}$, the induction equation (\ref{dimlessInd}) yields:
\begin{eqnarray}
\partial_t {\hbB}_{1/2} + R [({\bf e}_x + {\bf e}_y) \cdot \bnabla_{\bf x}] {\hbB_{1/2}} -  \bnabla_{\bf x}^2 {\hbB_{1/2}} &  = & (\cB(Z,T) \cdot \bnabla_{\bf x})  {\hbv}_{1/2}  \, . \label{orderhalf}
\end{eqnarray}
This is an equation for ${\hbB}_{1/2}$, with a forcing on the right-hand side. The solution is the sum of a particular solution, plus a solution to the homogeneous equation. The latter has exactly the same form as ${\hbB}_0$, because the linear operator is the same in both (\ref{order0}) and (\ref{orderhalf}). We can therefore include the solution to the homogeneous equation into $\cB(Z,T)$ and ask for the particular solution to have a vanishing average over $x$, $y$ and $t$: 
\begin{eqnarray}
\la \hbB_{1/2} \ra = \boldsymbol{0} \, .
\end{eqnarray}
To obtain the equation governing the evolution of $\cB$, we collect terms of order $\epsilon^2$ in the induction equation (\ref{dimlessInd}):
\begin{eqnarray}
\partial_t {\hbB}_{2} + \partial_T \cB + R [({\bf e}_x + {\bf e}_y) \cdot \bnabla_{\bf x}] {\hbB_{2}}   &  = & \bnabla_{\bf x} \times(\hbv_{1/2} \times \hbB_{3/2}) + \bnabla_{\bf X} \times(\hbv_{1} \times \cB) \\
\nonumber & & + \bnabla_{\bf X} \times(\hbv_{1/2} \times \hbB_{1/2})  +  \bnabla_{\bf x}^2 {\hbB_{2}} +  \bnabla_{\bf X}^2 {\cB}\, ,
\end{eqnarray}
before taking the average over $x$, $y$ and $t$:
\begin{eqnarray}
\partial_T \cB &  = &  \bnabla_{\bf X} \times ( \la \hbv_{1} \ra \times \cB )  + \bnabla_{\bf X} \times \la \hbv_{1/2} \times \hbB_{1/2} \ra  + \bnabla_{\bf X}^2 {\cB}\, .
\end{eqnarray}
We wish to show that the first term on the right-hand side vanishes. We expand it as: 
\begin{equation}
\bnabla_{\bf X} \times ( \la \hbv_{1} \ra \times \cB )  = \la \hbv_{1} \ra (\bnabla_{\bf X} \cdot \cB)    - (\la \hbv_{1} \ra \cdot \bnabla_{\bf X} )\cB + (\cB \cdot \bnabla_{\bf X} ) \la \hbv_{1} \ra - \cB (\bnabla_{\bf X} \cdot \la \hbv_{1} \ra) \, .
\end{equation}
The $x$ and $y$ average of the ${\cal O}(\epsilon^2)$ incompressibility constraint yields $\bnabla_{\bf X} \cdot  \la \hbv_{1} \ra = 0$, which, together with vertical momentum conservation, leads to $\la \hbv_{1} \ra \cdot {\bf e}_z = 0$. Hence $ \cB (\bnabla_{\bf X} \cdot \la \hbv_{1} \ra) = \boldsymbol{0}$ and $(\la \hbv_{1} \ra \cdot \bnabla_{\bf X} )\cB = \boldsymbol{0}$. The $x$ and $y$ average of the ${\cal O}(\epsilon)$ divergence-free constraint for ${\bf B}$
yields $\bnabla_{\bf X} \cdot  \cB = 0$. Because we do not allow for a pre-existing uniform background magnetic field in such a dynamo study, this leads to $\cB \cdot {\bf e}_z = 0$, and finally to $(\cB \cdot \bnabla_{\bf X} ) \la \hbv_{1} \ra  =\boldsymbol{0}$. We conclude that $\bnabla_{\bf X} \times ( \la \hbv_{1} \ra \times \cB ) = \boldsymbol{0}$, and we write the evolution equation for the large-scale magnetic field as:
\begin{eqnarray}
\partial_T \cB &  = & \bnabla_{\bf X} \times \la \hbv_{1/2} \times \hbB_{1/2} \ra  + \bnabla_{\bf X}^2 {\cB}\, . \label{SCorder2}
\end{eqnarray}

Finally, we need to include the equation governing the evolution of $\hbv_{1/2}$.Collecting terms of order $\epsilon^{1/2}$ in the Navier-Stokes equation (\ref{dimlessNS}), we obtain:

\begin{eqnarray}
\partial_t {\hbv}_{1/2} +   R [({\bf e}_x  + {\bf e}_y) \cdot \bnabla_{\bf x}] {\hbv}_{1/2} + & & \frac{R}{Ro} ({\bf e}_x + {\bf e}_y) \times {\hbv}_{1/2} \label{NSorderhalf}  \\
\nonumber & & = -\bnabla_{\bf x} \hat{p} + \Pm \, \bnabla_{\bf x}^2 {\hbv}_{1/2} + ({\cB} \cdot \bnabla_{\bf x}) {\hbB_{1/2}} + \hat{{\bf F}} \, , 
\end{eqnarray}
where we scaled the generalized pressure field as $p=\epsilon^{1/2} \hat{p}$. To lowest order, the divergence-free constraint becomes $\bnabla_{\bf x} \cdot \hbB_{1/2} = 0$.

In the main body of this study, we use the notations ${\bf b}=\epsilon^{1/2} \, \hbB_{1/2}$ and ${\bf v}=\epsilon^{1/2} \, {\hbv}_{1/2}$. Equations (\ref{orderhalf}), (\ref{SCorder2}) and (\ref{NSorderhalf}) then reduce to (\ref{ssind}), (\ref{lsind}) and (\ref{reducedNS}).

\section{Determination of the normal form \label{appnormalform}}
We consider the vicinity of the linear instability threshold and consider $\Rm=\Rm_c + \delta \, \Rm_1$, where $\delta \ll 1$. The domain is periodic in $z$ with spatial period $\lambda$. The magnetic field is expanded as:
\begin{eqnarray}
{\cal B}_{x;y}= \sqrt{\delta}\left( {\cal B}_{x;y}^{(0)}(t,T) + \delta {\cal B}_{x;y}^{(1)}(t,T) + \dots \right) \, ,
\end{eqnarray}
where we introduced the slow time $T=\delta \, t$. The dimensionless $\alpha$-effect coefficients are expanded as:
\begin{eqnarray}
{\alpha}_{xx;yy} & = & - \frac{\Rm}{1+R^2} \left(  1+\frac{2{\cal B}_{x;y}^2}{(1+R^2)(1-Ro^{-1})}   \right) \, .
\end{eqnarray}
To order $\sqrt{\delta}$, equations (\ref{eqBx}-\ref{eqBy}) lead to:
\begin{eqnarray}
\partial_{{t}} \Bx^{(0)} & = &  \frac{\Rm_c^2}{1+R^2} \, \partial_{{z}}  \By^{(0)} +  \partial_{{{z}} {{z}}} \Bx^{(0)} \, ,\\
\partial_{{t}} \By^{(0)} & = & - \frac{\Rm_c^2}{1+R^2} \, \partial_{{z}}   \Bx^{(0)} +  \partial_{{{z}}{{z}}} \By^{(0)} \, , 
\end{eqnarray}
and the only solution that does not rapidly decay to $0$ as $t$ increases is the marginally stable one:
\begin{eqnarray}
\Bx^{(0)} +i \By^{(0)} = A(T) \exp\left( i \frac{2 \pi \ell}{\lambda} z \right)\, .
\end{eqnarray}
At order $\delta^{3/2}$, we obtain:
\begin{eqnarray}
\partial_{{t}} \Bx^{(1)}-  \partial_{{{z}} {{z}}} \Bx^{(1)} - \frac{\Rm_c^2}{1+R^2} \, \partial_{{z}}  \By^{(1)} & = & \frac{2 \Rm_c \Rm_1}{1+R^2}  \partial_{{z}}  \By^{(0)} \\
\nonumber & & + \frac{2 \Rm_c^2}{(1+R^2)^2(1-Ro^{-1})}  \partial_{z} [ (\By^{(0)})^3 ] -\partial_T \Bx^{(0)} \, , \\
\partial_{{t}} \By^{(1)}-  \partial_{{{z}} {{z}}} \By^{(1)} + \frac{\Rm_c^2}{1+R^2} \, \partial_{{z}}  \Bx^{(1)} & = & - \frac{2 \Rm_c \Rm_1}{1+R^2}  \partial_{{z}}  \Bx^{(0)} \\
\nonumber & & - \frac{2 \Rm_c^2}{(1+R^2)^2(1-Ro^{-1})}  \partial_{z} [ (\Bx^{(0)})^3 ]  -\partial_T \By^{(0)}  \, . 
\end{eqnarray}
Adding $i$ times the second equation to the first one, we obtain:
\begin{eqnarray}
\left( \partial_{t} - \partial_{zz} + \frac{i \Rm_c^2}{1+R^2} \partial_z  \right) \left\{ \Bx^{(1)} +i \By^{(1)}   \right\} & = & \left( - \partial_{T} -  \frac{2 i \Rm_c \Rm_1}{1+R^2} \partial_z  \right)  \left\{ \Bx^{(0)} +i \By^{(0)}   \right\} \\
\nonumber & & - \frac{2 i \Rm_c^2}{(1+R^2)^2(1-Ro^{-1})} \partial_z[ (\Bx^{(0)})^3 + i (\By^{(0)})^3 ] \, .
\end{eqnarray}
The solvability condition is obtained by demanding that the right-hand side have no terms proportional to $\exp\left( i \frac{2 \pi \ell}{\lambda} z \right)$. After substituting $\Bx^{(0)}=[A(T) \exp\left( i \frac{2 \pi \ell}{\lambda} z \right) + \bar{A}(T) \exp\left( -i \frac{2 \pi \ell}{\lambda} z \right)]/2$ and $\By^{(0)}=[A(T) \exp\left( i \frac{2 \pi \ell}{\lambda} z \right) - \bar{A}(T) \exp\left( -i \frac{2 \pi \ell}{\lambda} z \right)]/(2i)$ and collecting the terms proportional to $\exp\left( i \frac{2 \pi \ell}{\lambda} z \right)$, we obtain the normal form (\ref{normalform}).

\section{Definitions of the elliptic integrals \label{appelliptic}}

The results of this study are presented using elliptic integrals written in Jacobi's form. The definitions of the incomplete elliptic integrals are:
\begin{eqnarray}
{\cal E} \left( x \, ; k \right) & = & \int_{0}^{x} \frac{\sqrt{1-k^2 t^2}}{\sqrt{1-t^2}} \mathrm{d}t \, , \\
{\cal F} \left( x \, ; k \right) & = & \int_{0}^{x} \frac{ \mathrm{d}t }{\sqrt{(1-t^2)(1-k^2 t^2)}}\, .
\end{eqnarray}



\section{Numerical simulations \label{app:Numsimuls}}

With the goal of confirming the analytical solutions presented in figure \ref{biffixedR}, we have performed numerical simulations of the full MHD equations (\ref{dimlessNS})-(\ref{dimlessInd}). The code uses a pseudo-spectral method with standard de-aliasing, the fields being decomposed on a Fourier basis in all three directions inside a domain $\left( 2 \pi \ell, 2 \pi \ell, \lambda \right)$. We use a semi-implicit second-order Runge-Kutta time-stepping scheme with adaptive time-step. After a transient, the simulations settle into a steady state. We extract $M=\max_z\{\la{\bf B}\ra \cdot {\bf e}_x\}/\sqrt{|1-Ro^{-1}|}$ in this steady state, and we plot $M$ as a function of $\Rm \sqrt{\lambda/\ell}$ in figure \ref{biffixedR}. The dimensionless parameters for the three different sets of runs shown in figure \ref{biffixedR} are:
\begin{itemize}
\item Case 1: $R = 0.5, Ro = 0.05, Re = 0.5, \Pm = 1, \lambda/\ell = 32 \pi $,
\item Case 2: $R = 2.0, Ro = -0.2, Re = 2.0, \Pm = 1, \lambda/\ell = 128 \pi $,
\item Case 3: $R = 2.0, Ro = 0.2, Re = 2.0, \Pm = 1, \lambda/\ell = 128 \pi $.  
\end{itemize}

\end{document}